\documentclass[preprint,11pt]{article}
\usepackage{setspace}
\usepackage{amssymb}
\usepackage{graphicx}
\usepackage{subcaption}
\usepackage{mathtools}
\usepackage{hyperref}

\usepackage{amsthm}
\newtheorem{definition}{Definition}

\newcommand*{\mc}[1]{\mathcal{#1}}

\usepackage{amsmath,amsthm,amssymb}
\newcommand\newsubcap[1]{\phantomcaption%
	\caption*{\figurename~\thefigure\thesubfigure: #1}}

\providecommand{\keywords}[1]{\textbf{\textit{Keywords---}} #1}

\begin{document}

\title{Deep Structure Learning using Feature Extraction in Trained Projection Space}


\author{Christoph Angermann and  Markus Haltmeier\\ \\ 
	Department of Mathematics, University of Innsbruck\\ Technikerstra{\ss}e 13, A-6020 Innsbruck\\
	\tt\{christoph.angermann,markus.haltmeier\}@uibk.ac.at\\
	\url{http://applied-math.uibk.ac.at}}

\maketitle

\begin{abstract}
Over the last decade of machine learning, convolutional neural networks  have been the most striking successes for feature extraction of rich sensory and high-dimensional data.  While learning data representations via convolutions is already well studied and efficiently implemented in various deep learning libraries, one often faces limited memory capacity and  insufficient number of training data, especially for high-dimensional and large-scale tasks. To overcome these limitations, we introduce a network architecture using a self-adjusting and data dependent version of the Radon-transform (linear data projection), also known as x-ray projection, to enable feature extraction via convolutions in lower-dimensional space. The resulting framework, named PiNet, can be trained end-to-end and shows promising performance on volumetric segmentation tasks. We  test proposed model on public datasets to show that our approach achieves comparable results only using fractional amount of parameters. Investigation of memory usage and processing time confirms PiNet's superior efficiency compared to other segmentation models.\\
\end{abstract}






\section{Introduction}
Deep convolutional networks have experienced  tremendous success in various scientific applications in the last few years \cite{segmentation, tian20}. An important contribution to the development of  network architectures was made in 2015 by Simonyan and Zisserman \cite{vgg}. The VGG Net redefined the state-of-the-art in image classification utilizing deep multi-scale feature learning via convolutional blocks with small $3\times 3$ filters. With this approach for feature extraction of rich-sensory data, the Visual Geometry Group of University of Oxford secured the first and the second places in the localisation and classification tracks of the ImageNet Challenge 2014 \cite{imagenet,ilsvrc}, respectively. In addition to VGG Net's superior capability of capturing image distribution, its generalization properties suggested extending this architecture beyond classification tasks. John Long et al. proposed in \cite{JL15} the idea of leveraging VGG Nets to deep convolutional filters to obtain semantic segmentations of 2D images. As an extension of the approach of fully convolutional filters, Ronneberger et al. \cite{OR15} introduced the U-net architecture, which redefines the benchmarks in image segmentation till today. The U-net provides a powerful segmentation and restoration tool, in particular for biomedical applications, due to its ability to learn highly accurate annotation masks from only few training data. 

%Although the U-net originally was considered as segmentation framework, employing this structure to other application tasks showed promising performance. 
%As an example, Hyun et al. \cite{mri} make use of the U-net for reconstruction of brain MRI scans out of undersampled $k$-space data. Since $k$-space subsampling in the time-consuming phase-encoding direction causes aliasing artefacts in image space, Hyun et al. showed that fully convolutional networks in combination with advanced subsampling techniques enables high quality image reconstruction as effectively as standard MRI reconstruction with the fully sampled data.

Feature extraction utilizing VGG Net architecture in the afore mentioned applications mainly focuses on 2D convolutions. Representation learning via convolutions in three-dimensional data space yields memory issues, significant increase of time consumption and often suffers from limited amount of available annotated training data \cite{CA19, reinvention,mpunet}. For segmentation, one idea to overcome the drawbacks of 3D convolutions is to apply parameter efficient 2D segmentation methods to slice images \cite{segmentation,OC16}. However, these slice images do not contain information of the full volumetric data which makes any task much more challenging.

Therefore, Perslev et al. \cite{mpunet} suggested to sample input volumes on 2D isotropic grids along multiple view axes. Different slice images are obtained for each axis, which can be used to train one 2D U-net implicitly enabling data augmentation. This yields different segmentation volumes for each view, which are combined again by a learned fusion model. Perslev et al. achieve remarkable performance on the 2018 Medical Segmentation Decathlon \cite{decathlon}, where they prove that the proposed resources saving approach can compete with much more complex methodologies on such generalization challenges. Nevertheless, with an underlying 2D segmentation model of about $ 62\times 10^6$ parameters and with a lot of redundant computations (slice-wise approach for each of these multiple view axes), there is still room for increasing time and memory efficiency.

Another attempt on memory reduction and data augmentation in the setting of high-dimensional segmentation was made by Angermann et al. in \cite{CA19}, where the Maximum Intensity Projection (MIP) in combination with a learnable reconstruction operator is used to transfer a 3D segmentation problem to the task of 2D segmentation of projection images. The choice of the MIP was motivated by the targeted application of blood vessel segmentation and is not directly applicable to annotation of arbitrary non convex 3D structures.\\

\textbf{Contributions:}
This paper presents a methodology for feature extraction of high-dimensional large-scale data using segmentation algorithms in a learned projection space. By developing a data-driven Radon-transform operator, the segmentation task is transferred to a bundle of regression problems in lower-dimensional space, which can be solved with vanilla U-net. This approach especially addresses two challenges of high-dimensional feature learning:
\begin{itemize}
	\item \textbf{Memory efficiency}: The need of large memory resources during training is reduced because representation learning happens via convolutions in a space with smaller dimension.
	\item \textbf{Implicit data augmentation:} Since we are deploying the Radon-transform from multiple directions,  we obtain several  projection images out of each input volume, which results into implicit data augmentation for the applied convolution algorithm in lower dimension. 
\end{itemize}
Furthermore, this paper proposes the following novelty:
\begin{itemize}
	\item \textbf{Data-driven Radon-transform:} For each projection direction, the Radon-transform operator \cite{radon} is combined with input dependent region weightings, which are adjusted automatically during  training process. This ensures catching as much as possible information of the volumetric object of interest while reducing artefacts caused by surroundings.
	
%	\item \textbf{Depth discretization:}
%	Utilizing the Radon-transform turns a classical classification problem into a bundle of regression tasks. Since well-known regression losses resulted in limited accuracy, we propose to transfer regression into a classification task again utilizing depth discretization which enables hierarchical segmentation of different height levels using standard classification loss functions. 
%	\item \textbf{Hierarchical convolution module:} Utilizing the Radon-transform turns a classical classification problem into a bundle of regression tasks. Since well-known regression losses (e.g. MSE, MSLE , MAE, etc.) resulted in limited accuracy, we introduce a novel hierarchical convolution module enabling  hierarchical segmentation of different height levels using standard classification loss functions. 
	
\end{itemize}
We test PiNet on selected data sets with only few training samples, taken from the 2018 Medical Segmentation Decathlon \cite{decathlon}. To be more precise, we apply exactly the same structure (without any task-specific adaptions) to two datasets of different modality (mono-modal MRI and CT), both with very small training data availability, and show that our approach is able to succeed on both differing tasks only using a fractional amount of parameters, memory and processing time compared to other challenge submissions, which also focused on storage efficiency \cite{mpunet}. Also the effect of steadily decreasing the amount of training pairs is investigated, where PiNet also succeeds on supervised training with marginal amount of available pairs. The numerical experiments are intended to demonstrate the universality and economy of projection based feature learning and also to propose the idea of switching from slice-wise to projection-based approaches when targeting high-dimensional image computations. Automated fine-tuning for particular applications is left for future research.

Section 2 gives a precise description of proposed network architecture, where every model component is explained in detail and necessity of each propagation step is justified. In Section 3, comprehensive evaluation studies on public challenge datasets are presented, where different parameter specifications of proposed architecture are evaluated and compared to benchmark models for volumetric segmentation. The paper is wrapped up by discussion on observed results.

\section{Method}

\begin{figure}[h!]
	\centering
	\includegraphics[width=.7 \textwidth]{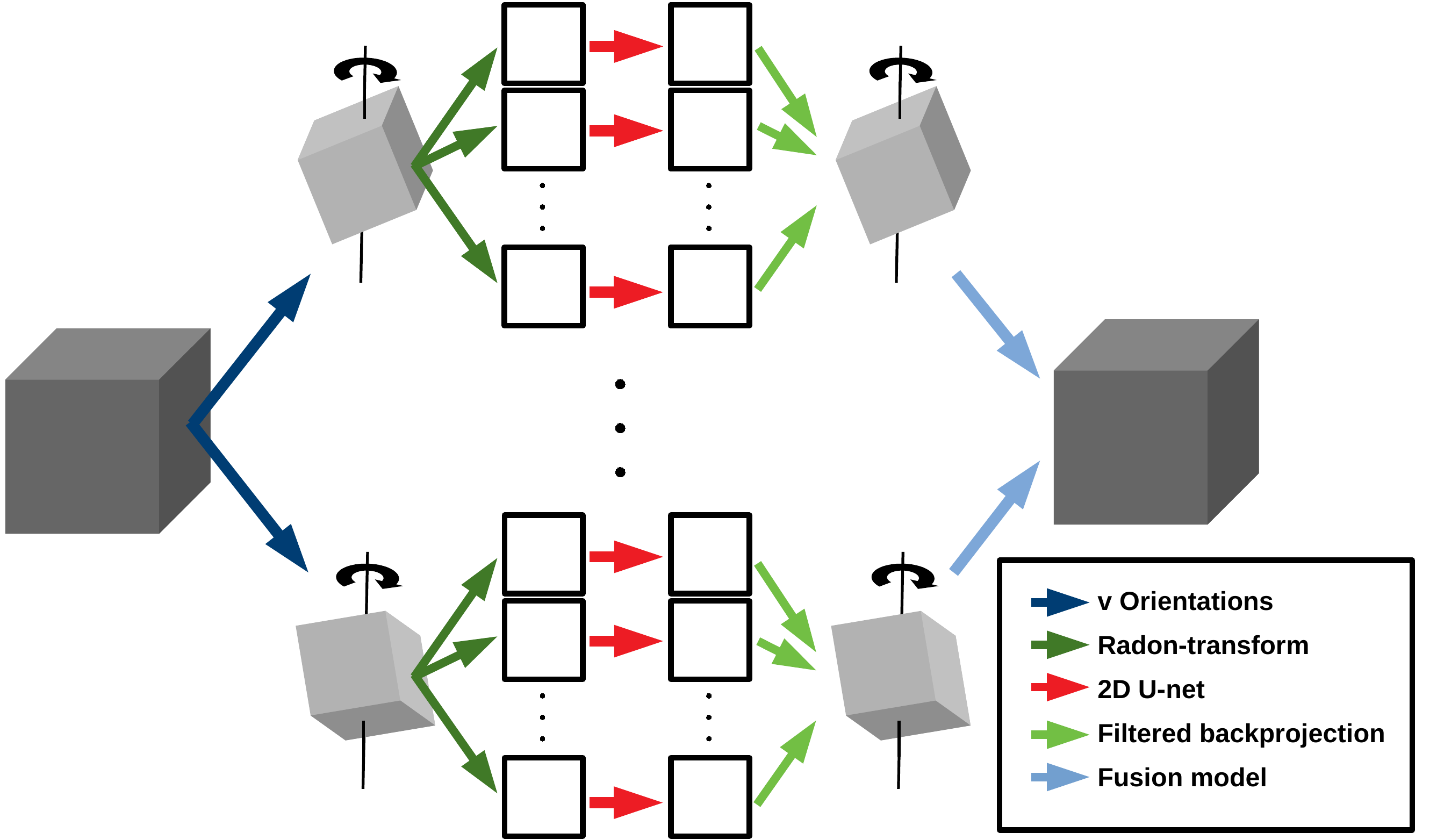}\hfil
	\includegraphics[width=.3 \textwidth]{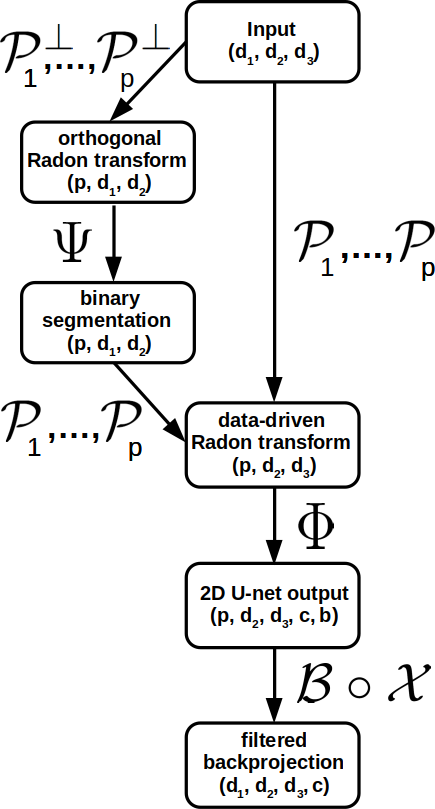}
	\caption{Left: Visualization of the PiNet architecture in Equation (\ref{eq1}). Right: The flowchart describes model forward propagation for one ground orientation ($v=1$).}
	\label{fig:1}
\end{figure}
In this section, we describe the proposed PiNet in three spatial dimensions and consider the task of volumetric segmentation. Note, that our approach also can be adapted to dimension reduction in spaces with arbitrary dimension.

Recall, that a network for multilabel volumetric classification is a function $\mathcal{N}\colon \mathbb{R}^{d_1\times d_2 \times d_3}\to [0,1]^{d_1 \times d_2\times d_3 \times c}$, that maps an input scan of size $d_1\times d_2 \times d_3$ to the probabilities that each voxel corresponds to one of the considered $c$ classes.
\begin{definition}[PiNet]
	For input $x\in\mathbb{R}^{d_1\times d_2\times d_3}$, the proposed PiNet (Figure \ref{fig:1}) takes the following form:
	\begin{align}\label{eq1}
	\Pi(x) = \mc{F} \circ \begin{bmatrix}
	\mc{B} \circ
	\begin{bmatrix}
	\mc X \circ \Phi \circ \mc P_1 \big( \mc{Q}_1 (x), \Psi \circ \mc P_1^\perp \circ \mc Q_1(x)\big)\\
	\vdots\\
	\mc X \circ \Phi \circ \mc P_p \big( \mc{Q}_1 (x), \Psi \circ \mc P_p^\perp \circ \mc Q_1(x)\big)
	\end{bmatrix}\\ \\
	 \bullet\\ \bullet\\\bullet\\ \\
	\mc{B} \circ
	\begin{bmatrix}
	\mc X \circ \Phi \circ \mc P_1 \big( \mc{Q}_v (x), \Psi \circ \mc P_1^\perp \circ \mc Q_v(x)\big)\\
	\vdots\\
	\mc X \circ \Phi \circ \mc P_p \big( \mc{Q}_v (x), \Psi \circ \mc P_p^\perp \circ \mc Q_v(x)\big)
	\end{bmatrix}\\
	\end{bmatrix}.
	\end{align}
	Different operators of Equation (\ref{eq1}) are discussed in the following paragraphs.
\end{definition}

\noindent
\textbf{Rotations $\mc Q_i$:}\\
$\mc Q_i: \mathbb{R} ^{d_1\times d_2 \times d_3}\to \mathbb{R}^{d_1 \times d_2\times d_3},\ i=1,\ldots,v$, denote rotations of the volumes in three-dimensional space, where $\mc Q_1$ denotes the identity, $\mc Q_2$ rotation of 90 degrees around the $x$-axis and $\mc Q_3$ rotation of 90 degrees around the $y$-axis. More precisely, we change the orientation of the input volumes corresponding to the the first two spatial axes to enable data augmentation for the subsequent models and to gain some multi-view information for the Radon-transform operator, where data only is rotated around the $z$-axis (Figure \ref{fig:1}). Therefore, $v=1$ corresponds to no additional orientations, $v=2$ denotes one additional and $v=3$ denotes two additional orientations. \\ \\
\textbf{Data-driven Radon-transforms $\mc P_i$:} \\
$\mc P_i:\mathbb R^{d_1 \times d_2\times d_3} \times \mathbb [0,1]^{d_1\times d_2} \to \mathbb{R}^{ d_2\times d_3},\ i=1,\ldots,p,$ is a Radon-transform  operator (projection along parallel lines), also known as x-ray transform, applied to input $\tilde x\in  \mathbb{R} ^{d_1\times d_2 \times d_3}$ and region weights $q \in [0,1] ^{d_1\times d_2}$, which enables reduction by one dimension. Basically, for $M\in \mathbb N$, the volume is rotated around the third coordinate axis for equidistant angles in
\begin{equation}\label{eq2}
\Theta \triangleq\bigg\{k\times \frac{180\cdot M}{d_1\cdot\pi} -90 \biggm| k=0,1,\ldots,\bigg\lfloor\frac{d_1\cdot \pi}{M}\bigg\rfloor-1\bigg\}
\end{equation}
and for each direction the rotated slices are summed over the first spatial axis. The quite complex notation of $\Theta$ in Equation (\ref{eq2}) indicates, that  for $M=2$ the Nyquist-Shannon sampling implies, that the volumetric data sample can be uniquely recovered from its Radon projections.

\begin{figure}[h!]
	\centering
	\includegraphics[width=0.95\textwidth]{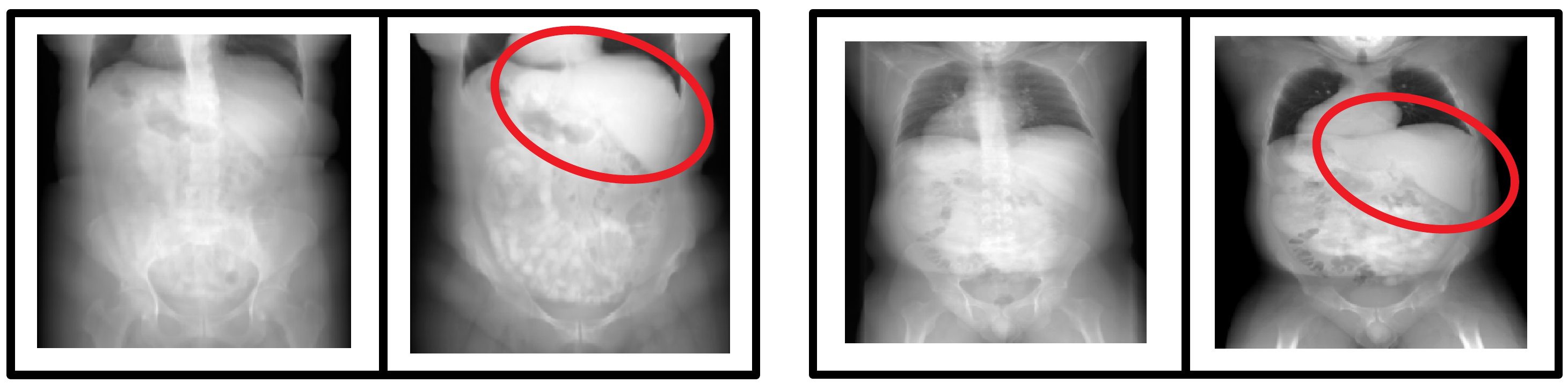}
	
	\caption{In both images, we see application of the Radon-transform to liver CT scans taken from the Medical Segmentation Decathlon \cite{decathlon}.  The left projection images are generated by integration over the first spatial axis without any region weighting along the projection direction (plain Radon-transform). The right images display the output of  data-driven Radon-transform operator, which pays more attention to regions along the projection axis containing liver. As a result, liver shape is better recognizable in the right projection images (red marked areas), which increases  accuracy for the subsequent segmentation model.}
	\label{fig:3}
\end{figure}

\begin{figure}[h!]
	\centering
	\includegraphics[width=0.89\textwidth]{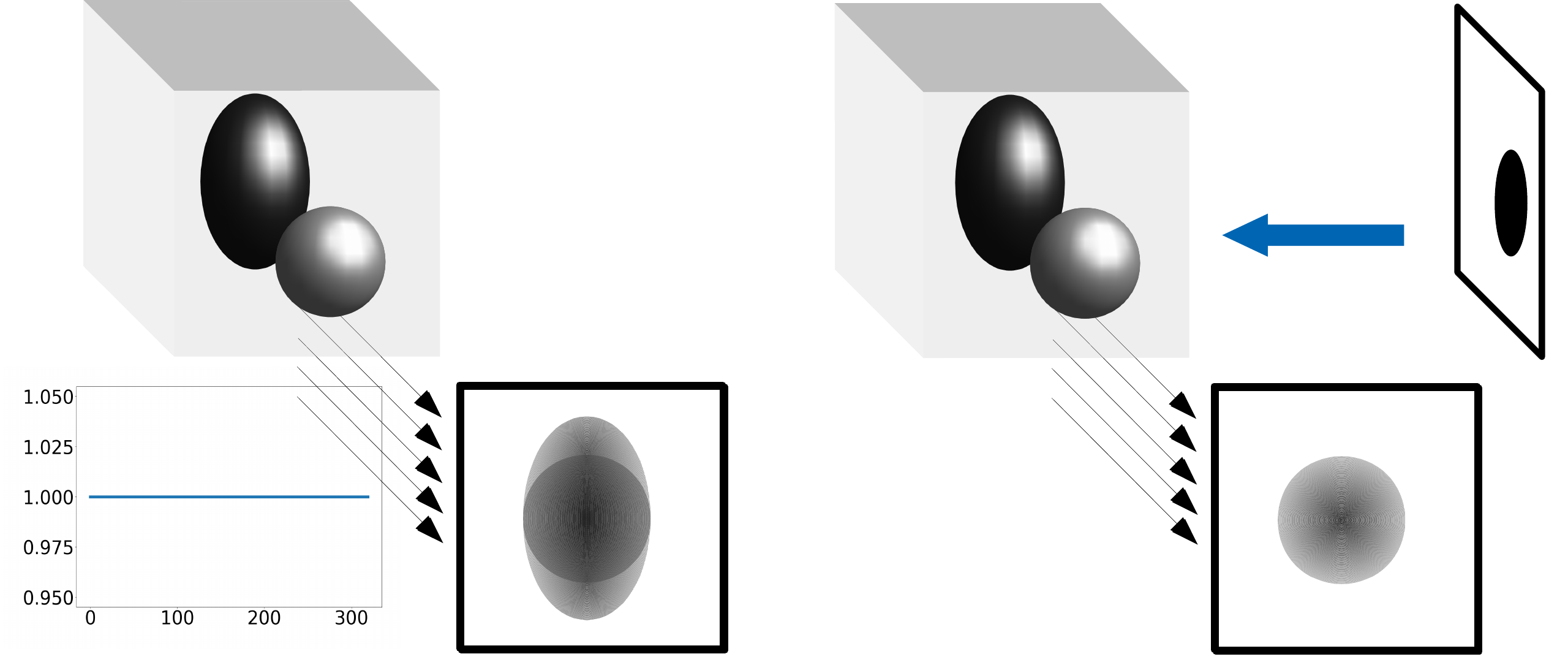}
	
	\caption{\textit{Left:} Vanilla Radon-transform integrating equally over both ellipsoids, where each slice has equal contribution.
		\textit{Right:} Multiplication with an automatically generated binary mask from the orthogonal direction ensures that the data-driven Radon-transform $\mc P$ pays more attention to those regions which contain the ellipsoid of interest.}
	\label{fig:2}
\end{figure}

When summing over parallel lines without weighting, we not only collect structural information of the targeted object we want to segment but also of surrounding objects. This causes loss of essential information like abnormalities (e.g. cancer) within the object and shades clear boundaries, which makes the segmentation in projection space much more challenging. As an example, segmentation of the liver is more difficult if another object like the vertebra contributes higher intensity as the targeted object to the projection image (Figure \ref{fig:3}). This issue can be resolved by automatically removing projection contributions from those objects, which surround the segmentation target. Since the position of the target object varies between different input scans, we propose an auxiliary binary segmentation from the orthogonal direction, which assists us determining the coarse position of the target in an automatic manner (Figure \ref{fig:2}).
\\

\noindent
\textbf{Auxiliary orthogonal position mask $\Psi \circ \mc P^\perp$:} \\
As illustrated in Figure \ref{fig:2}, 
$\mc P_i^\perp :\mathbb R^{d_1 \times d_2\times d_3}  \to \mathbb{R}^{ d_1\times d_2},\ i=1,\ldots,p$ is a standard Radon-transform  operator which generates projection images from the orthogonal direction $\alpha_i + 90,\ \alpha_i \in \Theta$ in Equation (\ref{eq2}).\\
$\Psi: \mathbb R^{d_1\times d_2}\to \mathbb [0,1]^{d_1\times d_2}$ is a simple U-net for binary segmentation of 2D projection images. For each channel, the network returns values near 1 if the integrated line of the projection image contained a voxel of one of the targeted objects and values near 0 else. The purpose of this mask from orthogonal direction is to assist the subsequent data-driven Radon-transform operator with a coarse position of the targets, which makes automated region emphasis in $\mc P$ input-dependent.
%This issue can be resolved by weighting the slices along the first spatial dimension for each position on second spatial axis before integration (Fig. \ref{fig:2}). During training of the subsequent segmentation network, the Radon-transform operator learns which areas along the projection axis are more important for the annotation task, individually for each direction in $\Theta$ and each height level. The weight matrix  $W_{\mc P}\in \mathbb R ^{p\times d_1\times d_2}$ of $\mc P$ is taken from a simple convolutional autoencoder, which gets optimized in parallel with the subsequent 2D segmentation network $\Phi$.
\\
 
\noindent
\textbf{2D network $\Phi$:}\\
$\Phi: \mathbb R^{d_2\times d_3}\to \mathbb [0,1]^{d_2\times d_3\times c \times b}$ is a U-net for segmentation of 2D projection images following the architecture in \cite{OR15}. This U-net operator is responsible for feature extraction to be able to succeed on the targeted segmentation task. Since we consider Radon-transform projection inputs $ x_R \in \mathbb R^{d_2\times d_3}$, it is also necessary to apply plain Radon-transform (applied without region weighting) to the corresponding targets. Therefore, the targets $ y_R \in \mathbb R^{d_2\times d_3 \times c}$ are not binary any more which changes the segmentation into a regression task for each output channel. This suggests modification of the output layer of U-net architecture in \cite{OR15}. Due to high sparsity of targeted objects, well-known regression losses as mean-squared-(logarithmic)-error (MSE/MSLE), mean-absolute-error (MAE) or Huber-loss failed on that task. We therefore adopt the idea of discretizing  the targets as it has been proposed in depth estimation works \cite{depthmap1,depthmap2}. To be more precise, we divide every target $ y_R \in \mathbb R^{d_2\times d_3 \times c}$ by its maximum, discretize for the scaled versions the interval $[0,1]$ into $b$ equidistant bins and receive the new targets $y_d \in \mathbb \{0,1\}^{d_2\times d_3 \times c \times b}$. The regression problem is therefore transformed again into a segmentation problem for each of the $b$ grey levels. 

For binary segmentation, deployment of convolution with softmax-activation-function as output layer
enables optimization via standard classification loss functions (cross-entropy-loss \cite{bishop}, Dice-loss \cite{vnet, dice}, etc.) separately on each depth channel while maintaining dependency between bins.\\ \\ 
\textbf{Filtered backprojection $\mc B\circ \mc X$:}\\
The segmented projection images $\hat y_d=[o_1,\ldots,o_b]\in \mathbb [0,1]^{d_2\times d_3 \times c \times b}$ are weighted and summed over the last axis, smoothed by a learnable 2D convolution $\mathcal{X}$ and lift again to volumetric data using filtered backprojection algorithm \cite{CA19,radon}. The procedure is followed by a learnable scaling and sigmoid activation to obtain clear decision boundaries. Following these steps results into volumetric segmentation masks for each orientation $\hat y_1,\ldots,\hat y_v \in [0,1]^{d_1 \times d_2\times d_3 \times c}$.\\\\
\textbf{Fusion model $\mc F$:}\\
\label{fusion}
As proposed in \cite{mpunet}, we make use of a fusion model $\mc F:\big([0,1]^{d_1\times d_2\times d_3\times c}\big)^v \to [0,1]^{d_1\times d_2\times d_3\times c}$ to combine the different orientations to one final output $\hat y$:
\begin{align}
\hat y=\frac{1}{v}\bigg[\sum_{l=1}^{v} \mc Q_l^{-1}\big(\hat y_l\big)\cdot W[l,k]\bigg]_{k=1}^c,
\end{align}
where the weight matrix $W \in \mathbb R^{v\times c}$ is adjusted during framework training. Application of a threshold to the final output $\hat y$ yields the desired segmentation mask.

\section{Experiments \& Results}

We apply PiNet to three different segmentation tasks (without any task-specific modifications) taken from the 2018 Medical Segmentation Decathlon  \cite{decathlon}. Task 1 and task 2 consist of left cardiac atrium segmentation of mono-modal MRI and spleen segmentation of CT scans, respectively.

All in all, the whole PiNet framework consists of $\approx 12.6\times 10^{6}$ adjustable parameters, which is only fifth of the parameter number of MPUnet \cite{mpunet} ($\approx 62\times 10^6$), which constitutes in this work the benchmark for efficient 3D segmentation. Regarding Equation (\ref{eq1}), only $\approx 8\times 10^2$ parameters are used by the trainable operators $\mc X,\mc B,\mc F$, the rest is part of 2D networks $\Phi$ and $\Psi$. We choose  $v=1$ for the number of orientations (only the original input orientation), $b=5$ in the hierarchical output module and $M=32$ in Equation (\ref{eq2}) (i.e. 31  projection directions), since for higher choice only computational costs but not accuracy increase (cf. Table \ref{tab1}). Optimization is done enabling Dice-loss function \cite{vnet,dice} with Adadelta-algorithm \cite{adadelta} (learning rate = 0.2) and stopped after 130 epochs with minibatch of 10 projections per step. Everything is implemented in Python using the neural networks API Keras \cite{keras}. The parameters of $\Phi$ and $\Psi $ are initialized with the default Keras initialization, the start weights of $\mc X,\mc B,\mc F$ are initialized empirically.

We present results for PiNet via 3-fold cross validation on the available training pairs. Table \ref{tab1} shows perfomance for different modifications of PiNet architecture, Table \ref{tab2} displays for each task the corresponding amount of given training pairs and compares the Dice scores and Haussdorf distances (H-dist) \cite{hausdorff} in voxels between MPUnet \cite{mpunet}, 3D U-Net \cite{OC16,vnet, skull} and our methodology. Every framework in Equation Table \ref{tab2} \& \ref{tab1} is trained and evaluated on a standard 8GB \textit{NVIDIA GeForce RTX 2080} GPU. Last columns indicate how much GPU memory in Gigabyte is at least necessary to handle training with very small minibatches and the average processing time during application of each model.

For MPUnet, we use the GitHub-implementation (\url{https://github.com/perslev/MultiPlanarUNet}) with pre-defined parameter settings , since  Perslev et al. \cite{mpunet} claim that the fixed hyperparameter set ensures high generalizability and no further fine-tuning has to be conducted. For the 3D U-net approach with convolutions in three dimensions, we make use of a pre-built implementation in the same GitHub repository. Furthermore, we have to decrease input dimension from 320 to 80 for 3D U-net to ensure that training is still feasible for our GPU.
\begin{table}[htb!]
	\centering
	\caption{Comparison between MPUnet \cite{mpunet}, PiNet and an 3D U-net implementation \cite{OC16,vnet,skull} for selected datasets of the 2018 Medical Segmentation Decathlon \cite{decathlon}. For each task and network the reported values are Dice score and Hausdorff distance on unseen data (3-fold cross validation), minimal GPU memory in Gigabyte needed to train these implementations and average processing time during application in seconds.}
	\small
	\begin{tabular}{l || r|r |r  |r|r}
		\hline
		\textbf{Task/Size}&\textbf{Net}&\textbf{Dice score}&\textbf{H-dist.}&\textbf{GPU}&\textbf{Time} \\
		\hline
		Atrium/20& MPUnet&$0.911\pm 0.032
$& $\textbf{14.76}\pm 5.568
$ & $>5.6$ GB&169 sec \\ \hline
		
		&  3D U-net& $\textbf{0.918}\pm 0.029$&$21.27\pm32.20$& $>8.5$ GB&95 sec\\ \hline
		&  PiNet& $0.894\pm 0.037$&${15.76} \pm 5.455$& $>\textbf{3.3}$ GB&\textbf{22.6} sec\\ \hline
		Spleen/41& MPUnet &$0.923\pm 0.108$&$26.38\pm 42.93$& $>5.6$ GB&544 sec\\ \hline
		 
		&  3D U-net&$0.621\pm0.179 $&$265.5\pm84.13 $& $>8.5$ GB&492 sec\\ \hline
		& PiNet&$\textbf{0.937}\pm 0.040$ & $\textbf{9.650}\pm 6.191$& $>\textbf{3.3}$ GB&\textbf{23.6} sec\\ \hline
	\end{tabular}
	\label{tab2}
\end{table}
\begin{table}[htb!]
	\centering
	\caption{Performance of PiNet regarding number of directions $p$, number of orientations $v$ and weighted Radon-transform operator $\mc P$. Note that $v=1$ denotes use of only the original orientation, whereas $v=3$ denotes utilizing also the additional orientations obtained by rotating the input volume around the first two spatial axes.}
	
	\begin{tabular}{l || r|r |r  |r|r}
		\hline
		\textbf{Task/Size}&\textbf{p}&\textbf{v}&\textbf{$\mc P$ weighted} &\textbf{Dice score}&\textbf{H-dist.}\\
		\hline
		Atrium/20 & 15 & 1 & Yes & $ 0.883\pm 0.057
	$ & $22.77\pm 22.74$ \\  \hline
		& 31 & 1 & Yes & $0.894\pm 0.037$ & $15.76\pm 5.455$ \\  \hline
		& 62 & 1 & Yes & $0.890\pm0.048 $ & $16.47\pm5.701$ \\  \hline
		& 31 & 3 & Yes & $0.897\pm 0.043$ & $17.46\pm7.140 $ \\  \hline

		& 41 & 1 & No & $0.746\pm0.148$ & $21.66\pm7.560$ \\  \hline
		
		Spleen/41 & 15 & 1 & Yes & $0.924\pm 0.048$ & $11.08\pm 7.994$ \\  \hline
		& 31 & 1 & Yes & $0.937\pm 0.040$ & $9.650\pm 6.191$ \\  \hline
		& 62 & 1 & Yes & $0.932\pm0.038$ & $11.37\pm7.453$ \\  \hline
		
		& 31 & 3 & Yes & $0.945\pm0.025 $ & $9.514\pm5.026  $  \\  \hline
		& 41 & 1 & No & $0.726\pm0.154$ & $53.20\pm63.21$ \\  \hline
	\end{tabular}
	\label{tab1}
\end{table}
\begin{figure}[htb!]
	\centering
	\includegraphics[width=0.23\textwidth]{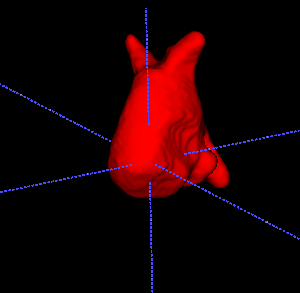}\hfil
	\includegraphics[width=0.23\textwidth]{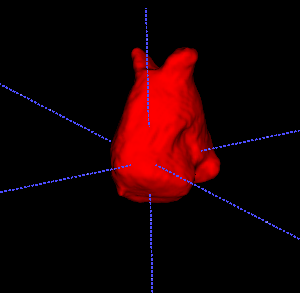}\hfil
	\includegraphics[width=0.23\textwidth]{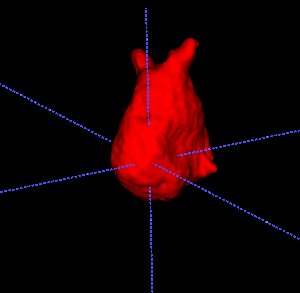}\hfil
	\includegraphics[width=0.23\textwidth]{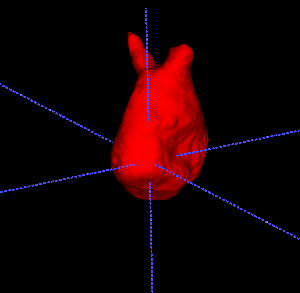}\vspace{0.25cm}
	\includegraphics[width=0.23\textwidth]{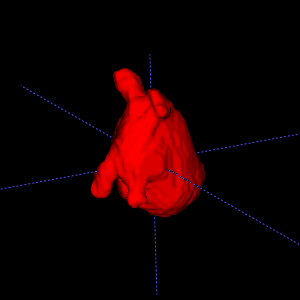}\hfil
	\includegraphics[width=0.23\textwidth]{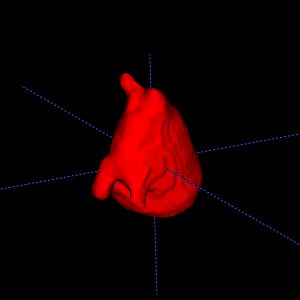}\hfil
	\includegraphics[width=0.23\textwidth]{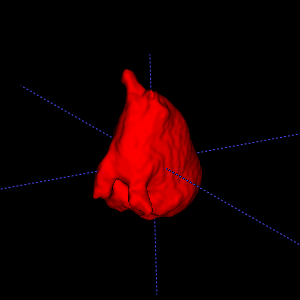}\hfil
	\includegraphics[width=0.23\textwidth]{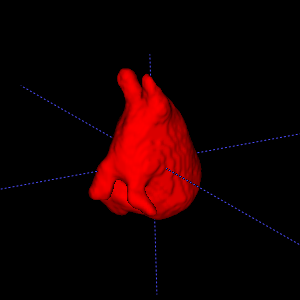}
	
	\newsubcap{From left to right: Left cardiac atrium annotations generated by hand (ground truth), PiNet, MPUnet, 3D U-net.}
	\label{fig:fig5}
\end{figure}
\begin{figure}[htb!]
	\centering
	\includegraphics[width=0.23\textwidth]{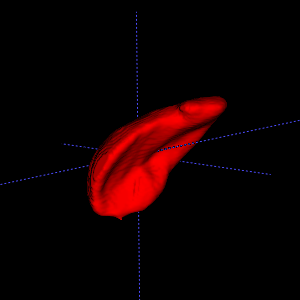}\hfil
	\includegraphics[width=0.23\textwidth]{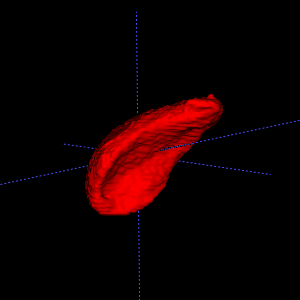}\hfil
	\includegraphics[width=0.23\textwidth]{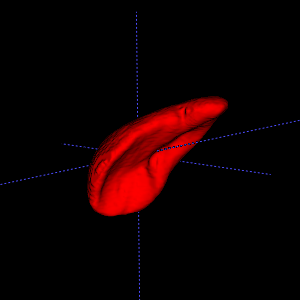}\hfil
	\includegraphics[width=0.23\textwidth]{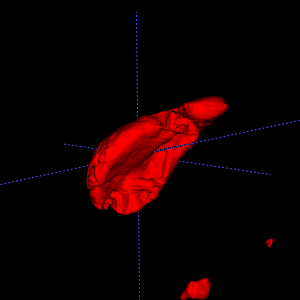}\vspace{0.25cm}
	\includegraphics[width=0.23\textwidth]{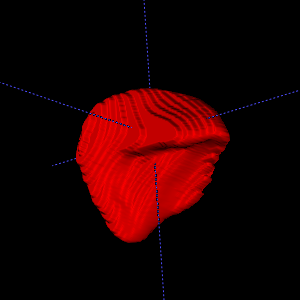}\hfil
	\includegraphics[width=0.23\textwidth]{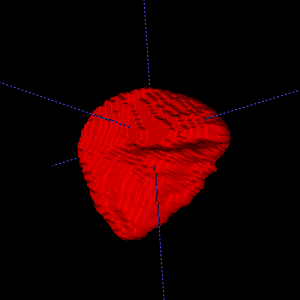}\hfil
	\includegraphics[width=0.23\textwidth]{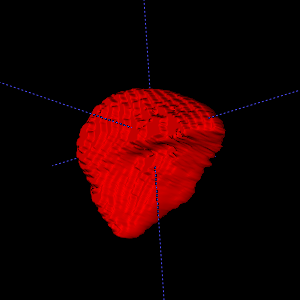}\hfil
	\includegraphics[width=0.23\textwidth]{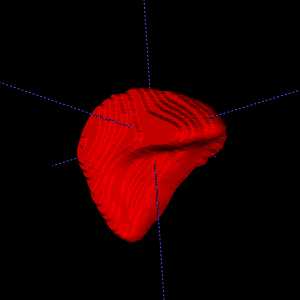}
	
	\newsubcap{From left to right: Spleen annotations generated by hand (ground truth), PiNet, MPUnet, 3D U-net.}
	\label{fig:fig6}
\end{figure}

As we observe in Table \ref{tab2}, PiNet, although it only needs fifth of the parameters and significantly less GPU memory, can compete with MPUnet, the considered benchmark for memory-efficient volumetric segmentation. PiNet achieves slightly smaller Dice scores for left cardiac atrium MRI segmentation, but significant higher Dice and decrease of average Hausdorff distance are reached for spleen CT annotation compared to MPUnet and 3D U-net. Exploring deviation values between all three methodologies we conclude, that there exist a few spleen CT scans in the test data where MPUnet and 3D U-net heavily fail. This could be caused by strong variations of spleen position and size in the very small train data set ($41\times \frac{2}{3}$ samples). Therefore, although PiNet is not able to significantly improve segmentation accuracy of very detailed structures like fine heart arteries (see Figure \ref{fig:fig5}), our approach exceeds other existing 3D segmentation methods in terms of capturing objects with strongly varying position and size in small data sets due to deploying multiple projection directions. 

Furthermore, PiNet significantly outperforms its comparable methodologies in terms of processing time during application, which makes our approach also quite relevant for deployment in field. Also necessity of transferring the Radon-transform to a trainable operator, which optimizes input-dependent region weightings, is confirmed by the test results in Table \ref{tab1}. Same table also suggests, that utilizing more than one ground orientation of the input volume ($v>1$) via Fusion model $\mc F$ (Equation \ref{fusion}) only results into a very slight improvement for both tasks in terms of evaluation metrics. From this observation we conclude, that in the two considered tasks the data-driven Radon-transform projection images of the ground orientation already contain sufficient 3D information of the input sample. It is also important to mention, that segmentation on Radon-transform projection images succeeds on both modalities MRT and CT, which are commonly used in biomedical applications.\\

\begin{figure}[h!]
	\centering
	\includegraphics[width=0.8\textwidth]{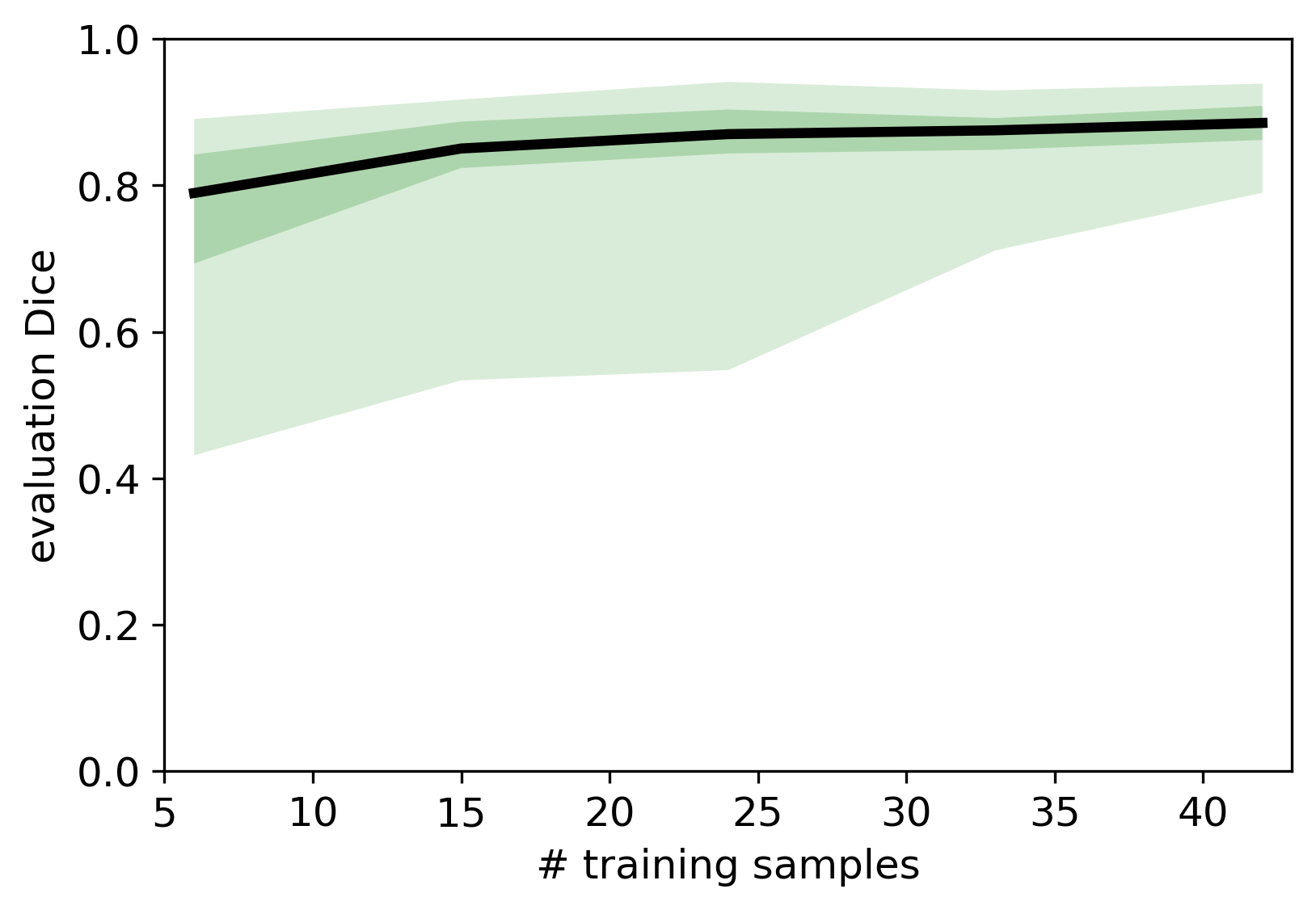}
	\caption{Statistical values for PiNet evaluation on unseen liver samples for varying amount of available training pairs and without data augmentation techniques. The plotted values are minimum, 1st quartile, median (black line), 3rd quartile and maximum.}
	\label{fig111}
	
\end{figure}

To further highlight the effect of implicit data augmentation for the projection-based PiNet approach, we choose a third dataset of the 2018 Medical Segmentation Decathlon \cite{decathlon}, which consists of 131 portal venous phase CT liver scans and their corresponding segmentations. We randomly select 44 of these liver pairs as our evaluation dataset and use small selections of the remaining 87 pairs as training data for our PiNet model. We start with a very small training dataset of size 6 and steadily increase the amount of training pairs by 9. For every training run, Dice metrics are reported for each amount of training samples (Figure \ref{fig111}). Only utilizing an amount of 6 training samples ensures reasonable performance on the 44 unseen evaluation cases without any data augmentation techniques. Although there exists a sample in the evaluation set where the minimal trained PiNet fails (Dice $< 0.5$), a median of nearly 0.8 is a satisfying result which steadily increases the more train data is added. Direct comparison to MPUnet and 3D Unet models in Figure \ref{fig111} is not possible since these implementations already include an automatic data augmentation preprocessing pipeline.\\

We were able to train a fast and efficient volumetric segmentation model, which needs GPU-resources of only 3.3 Gigabytes during optimization and generates satisfying annotations in a quite satisfying time on very small train data sets without any task-specific fine-tuning. This encourages us to further improve our PiNet implementation and, as a next step, to leverage it to a multilabel classification tool, so it can be easily applied to arbitrary volumetric segmentation tasks. 

\section{Conclusion}
To sum up, this work proposes a network architecture enabling a data-driven Radon-transform operator to transfer the segmentation task to lower-dimensional space. The segmentation task of projection imagles can be addressed with efficient model architectures (e.g. U-net) in lower-dimensional space, which we lift to original dimension deploying self-adjusting reconstruction operators. Application to datasets of the 2018 Medical Segmentation Decathlon with small train data availability encourage  universality and plausibility of our fast and memory-efficient PiNet approach. Applying the proposed PiNet on a third dataset with varying  amount of available training pairs suggests, that our approach reaches outstanding performance on very small data sets without the need for additional data augmentation implementations. Further steps include leveraging our approach to multilabel segmentation, and implementation of a fully-automated  self-tuning model framework, which can easily be applied to arbitrary high-dimensional segmentation tasks without need of large memory resources.



\bibliographystyle{ieeetr} 
\bibliography{ref}

\begin{thebibliography}{10}

\bibitem{segmentation}
N.~Tajbakhsh, L.~Jeyaseelan, Q.~Li, J.~N. Chiang, Z.~Wu, and X.~Ding,
  ``Embracing imperfect datasets: A review of deep learning solutions for
  medical image segmentation,'' {\em Medical Image Analysis}, p.~101693, 2020.

\bibitem{tian20}
C.~Tian, L.~Fei, W.~Zheng, Y.~Xu, W.~Zuo, and C.-W. Lin, ``Deep learning on
  image denoising: An overview,'' {\em Neural Networks}, 2020.

\bibitem{vgg}
K.~Simonyan and A.~Zisserman, ``Very deep convolutional networks for
  large-scale image recognition,'' in {\em ICLR}, 2015.

\bibitem{imagenet}
J.~Deng, W.~Dong, R.~Socher, L.-J. Li, K.~Li, and L.~Fei-Fei, ``Imagenet: A
  large-scale hierarchical image database,'' in {\em 2009 IEEE conference on
  computer vision and pattern recognition}, pp.~248--255, Ieee, 2009.

\bibitem{ilsvrc}
O.~Russakovsky, J.~Deng, H.~Su, J.~Krause, S.~Satheesh, S.~Ma, Z.~Huang,
  A.~Karpathy, A.~Khosla, M.~Bernstein, A.~C. Berg, and L.~Fei-Fei, ``{ImageNet
  Large Scale Visual Recognition Challenge},'' {\em International Journal of
  Computer Vision (IJCV)}, vol.~115, no.~3, pp.~211--252, 2015.

\bibitem{JL15}
J.~Long, E.~Shelhamer, and T.~Darrell, ``Fully convolutional networks for
  semantic segmentation,'' in {\em Proceedings of the IEEE conference on
  computer vision and pattern recognition}, pp.~3431--3440, 2015.

\bibitem{OR15}
O.~Ronneberger, P.~Fischer, and T.~Brox, ``U-net: Convolutional networks for
  biomedical image segmentation,'' in {\em International Conference on Medical
  image computing and computer-assisted intervention}, pp.~234--241, Springer,
  2015.

\bibitem{CA19}
C.~{Angermann}, M.~{Haltmeier}, R.~{Steiger}, S.~{Pereverzyev}, and
  E.~{Gizewski}, ``Projection-based 2.5d {U}-net architecture for fast
  volumetric segmentation,'' in {\em 2019 13th International conference on
  Sampling Theory and Applications (SampTA)}, pp.~1--5, 2019.

\bibitem{reinvention}
J.~Yang, X.~Huang, B.~Ni, J.~Xu, C.~Yang, and G.~Xu, ``Reinventing 2d
  convolutions for 3d medical images,'' {\em arXiv preprint arXiv:1911.10477},
  2019.

\bibitem{mpunet}
M.~Perslev, E.~B. Dam, A.~Pai, and C.~Igel, ``One network to segment them all:
  A general, lightweight system for accurate 3d medical image segmentation,''
  in {\em International Conference on Medical Image Computing and
  Computer-Assisted Intervention}, pp.~30--38, Springer, 2019.

\bibitem{OC16}
{\"O}.~{\c{C}}i{\c{c}}ek, A.~Abdulkadir, S.~S. Lienkamp, T.~Brox, and
  O.~Ronneberger, ``{3D} {U}-net: learning dense volumetric segmentation from
  sparse annotation,'' in {\em International conference on medical image
  computing and computer-assisted intervention}, pp.~424--432, Springer, 2016.

\bibitem{decathlon}
A.~L. Simpson, M.~Antonelli, S.~Bakas, M.~Bilello, K.~Farahani,
  B.~Van~Ginneken, A.~Kopp-Schneider, B.~A. Landman, G.~Litjens, B.~Menze, {\em
  et~al.}, ``A large annotated medical image dataset for the development and
  evaluation of segmentation algorithms,'' {\em arXiv preprint
  arXiv:1902.09063}, 2019.

\bibitem{radon}
S.~R. Deans, {\em The Radon transform and some of its applications}.
\newblock Courier Corporation, 2007.

\bibitem{depthmap1}
D.~Eigen, C.~Puhrsch, and R.~Fergus, ``Depth map prediction from a single image
  using a multi-scale deep network,'' in {\em Advances in neural information
  processing systems}, pp.~2366--2374, 2014.

\bibitem{depthmap2}
C.~Zhao, Q.~Sun, C.~Zhang, Y.~Tang, and F.~Qian, ``Monocular depth estimation
  based on deep learning: An overview,'' {\em Science China Technological
  Sciences}, pp.~1--16, 2020.

\bibitem{bishop}
C.~M. Bishop {\em et~al.}, {\em Neural networks for pattern recognition}.
\newblock Oxford University Press, 1995.

\bibitem{vnet}
F.~Milletari, N.~Navab, and S.-A. Ahmadi, ``V-net: Fully convolutional neural
  networks for volumetric medical image segmentation,'' in {\em 2016 Fourth
  International Conference on 3D Vision (3DV)}, pp.~565--571, IEEE, 2016.

\bibitem{dice}
T.~{Eelbode}, J.~{Bertels}, M.~{Berman}, D.~{Vandermeulen}, F.~{Maes},
  R.~{Bisschops}, and M.~B. {Blaschko}, ``Optimization for medical image
  segmentation: Theory and practice when evaluating with dice score or jaccard
  index,'' {\em IEEE Transactions on Medical Imaging}, vol.~39, no.~11,
  pp.~3679--3690, 2020.

\bibitem{adadelta}
M.~D. Zeiler, ``Adadelta: an adaptive learning rate method,'' {\em arXiv
  preprint arXiv:1212.5701}, 2012.

\bibitem{keras}
F.~Chollet {\em et~al.}, ``Keras.'' \url{https://keras.io}, 2015.

\bibitem{hausdorff}
D.~{Karimi} and S.~E. {Salcudean}, ``Reducing the {H}ausdorff distance in
  medical image segmentation with convolutional neural networks,'' {\em IEEE
  Transactions on Medical Imaging}, vol.~39, no.~2, pp.~499--513, 2020.

\bibitem{skull}
H.~Hwang, H.~Z.~U. Rehman, and S.~Lee, ``3d u-net for skull stripping in brain
  mri,'' {\em Applied Sciences}, vol.~9, no.~3, p.~569, 2019.

\end{thebibliography}

%\bigskip \bigskip
%
%\textbf{Christoph Angermann} received his MSc degree in mathematics in 2019 for his work on projection networks in biomedical volumetric segmentation. Since then he is PhD student at the Department of Mathematics, University of Innsbruck, Austria. His current research interests include deep learning, mathematical aspects of generative networks and self-supervised network training architectures.\\
%
%\textbf{Markus Haltmeier }received his PhD degree in mathematics from the University of Innsbruck, Austria, in 2007, for research on computed tomography.
%% He was then involved in various aspects of inverse problems as a research scientist with the University of Innsbruck, the University of Vienna, Austria, and the Max Planck Institute for Biophysical Chemistry, Göttingen, Germany. 
%Since 2012, he is full professor with the Department of Mathematics, University of Innsbruck. His current research interests include inverse problems, regularization theory, signal and image processing, computerized tomography, photoacoustic imaging and machine learning.

\end{document}